\documentclass[amsfonts,twocolumn,aps,prl]{revtex4}
\usepackage{graphicx}
\begin{document}
\title{Equilibrium magnetization at the boundary of a magnetoelectric antiferromagnet}

\author{K. D. Belashchenko}
\affiliation{Department of Physics and Astronomy and Nebraska Center of
Materials and Nanoscience, University of Nebraska-Lincoln, Lincoln, Nebraska
68588, USA}
\date{\today}

\begin{abstract}
Symmetry arguments are used to show that a boundary of a magnetoelectric
antiferromagnet has an equilibrium magnetization. This magnetization is
coupled to the bulk antiferromagnetic order parameter and can be switched
along with it by a combination of $\mathbf{E}$ and $\mathbf{B}$ fields. As a
result, the antiferromagnetic domain state of a magnetoelectric can be used
as a non-volatile switchable state variable in nanoelectronic device
applications. Mechanisms affecting the boundary magnetization and its
temperature dependence are classified. The boundary magnetization can be
especially large if the boundary breaks the equivalence of the
antiferromagnetic sublattices.
\end{abstract}

\maketitle

Magnetoelectric antiferromagnets (AFM) develop a magnetization $\mathbf{M}$
(or electric polarization $\mathbf{P}$) in the bulk when an electric (or
magnetic) field is applied \cite{LL8,Fiebig,Schmid}. This property is due to
the presence of a magnetoelectric term in the free energy,
$F_{\mathrm{ME}}=-\alpha_{ik}E_iH_k$, where $\alpha_{ik}$ is the
magnetoelectric tensor; the latter is odd under time reversal. An AFM is
magnetoelectric if the presence of an invariant polar vector $\mathbf{E}$ can
reduce its magnetic point group to a ferromagnetic one
\cite{Dzyal-Cr2O3,Schmid}.

Magnetoelectric and multiferroic materials can provide the necessary response
to allow electrical switching of the magnetic state
\cite{Fiebig,Eerenstein,RS,Cheong} and potentially enable fast, high-density,
low-power, and non-volatile memory devices (magnetoelectric memory)
\cite{BD,Bibes,Ramesh,Chu}. To enable easy readout of the magnetic state, the
magnetoelectric or multiferroic layer needs to be coupled to a proximate
ferromagnetic layer. This coupling requires an exchange bias
\cite{Meiklejohn,Nogues1,Nogues2} at the interface, which is the
time-reversal-breaking shift the hysteresis loop of the ferromagnet along the
magnetic field axis. Much attention in this context has been focused on the
room-temperature multiferroic BiFeO$_3$, but the destabilizing effects of
ferroelastic strains and depolarizing fields need to be circumvented for
non-volatile operation \cite{Baek}. Ferroelastic strain could be avoided by
using a multiferroic material with linear coupling of $\mathbf{P}$ and
$\mathbf{M}$, but suitable materials for room-temperature operation are not
available \cite{Fennie}.

An alternative approach to electric magnetization control uses the AFM order
parameter of a magnetoelectric material as the switchable state variable.
Magnetoelectric switching of Cr$_2$O$_3$ was shown \cite{Binek} to induce a
reversible switching of the exchange bias polarity in the proximate
ferromagnetic Pd/Co multilayer on the macroscopic scale. It was argued
\cite{Binek} that this effect is a manifestation of the equilibrium boundary
magnetization of a magnetoelectric, which is required by symmetry and couples
to the bulk AFM order parameter. Essentially, the boundary reduces the
symmetry in a similar way to the electric field. Another manifestation of
this effect is the spin polarization of the photoelectron current emitted
from the free Cr$_2$O$_3$ (0001) surface \cite{Binek}.

Macroscopic signatures of boundary magnetization of Cr$_2$O$_3$ \cite{Binek}
show that the lack of macroscopic time-reversal symmetry in a magnetoelectric
can translate into strong spin asymmetry at its boundary. However, the
microscopic mechanisms of this effect are not understood. In this Letter the
salient features of boundary magnetism of magnetoelectrics are analyzed from
the general point of view. A rigorous microscopic proof of the existence of
equilibrium boundary magnetization is given, and its microscopic mechanisms
are classified. In particular, it is shown that the effects can be very large
if the boundary breaks the equivalence of the AFM sublattices.

Consider a \emph{macroscopically} flat boundary (surface or interface) of an
AFM with an external normal $\mathbf{n}$, which is allowed to have roughness
and all possible terminations distributed with equilibrium Gibbs
probabilities. The magnetic structure of the boundary is also assumed to be
equilibrium, subject to the constraint that the bulk of the crystal is in the
single AFM domain state \cite{note-staggered}.

We are generally interested in the response of the boundary on the
macroscopic scale to an external probe which couples to the magnetic moment
density $\mathbf{m}(\mathbf{r})$. This probe can represent spin-resolved
photoemission, magneto-optic Kerr effect, exchange bias with a ferromagnet,
or magnetometry. For typical probes the measured quantity is an \emph{odd}
functional $G\{m_i(\mathbf{r})\}$, such that
$G\{m_i(\mathbf{r})\}=G\{m_i(\mathbf{r}+\mathbf{t})\}$ for any shift
$\mathbf{t}$ (or at least for any translation vector of the bulk lattice
treated as non-magnetic, such that $\mathbf{t}\cdot\mathbf{n}$ is not large
compared to the equilibrium roughness). Hereafter a probe is assumed to
satisfy this condition \cite{probe-note}. The component $m_i$ is selected by
the polarization of the probe.

The following arguments do not depend on the nature of the probe. For
definiteness, let us select the \emph{boundary moment} $\vec\mathcal{M}$ as
the probe, defined in a way that satisfies the above requirement of
translational invariance. Specifically, if the magnetic unit cell is larger
than the structural unit cell, the magnetic moment $\vec\mathcal{M}$ of the
boundary region must be averaged over the different ways of separating this
region from the bulk, related to each other by purely structural translations
(see Fig.\ 1). Surface-sensitive probes like exchange bias or spin-polarized
spectroscopies are free from this complication.

The macroscopic boundary magnetization is given by the equilibrium Gibbs
average $\mathbf{M}_{b}=\langle \vec\mathcal{M} \rangle/A$, where $A$ is the
boundary area, and the thermodynamic limit of large $A$ is assumed. Its
$i$-th component vanishes only if any boundary configuration (termination and
magnetic structure) has an energetically degenerate one with a reversed
$\mathcal{M}_i$; otherwise it is finite. Such degeneracy occurs only if the
bulk magnetic \emph{space} group of the crystal contains an operation under
which the vector $\mathbf{n}$ is invariant, and $\mathcal{M}_i$ is odd. All
degenerate boundary configurations can be generated by bulk space group
operations; this is the boundary generalization of the Aizu procedure
\cite{Aizu}. In particular, energetically degenerate atomic steps are
automatically accounted for, as shown in Fig.\ 1 and 2b. However, since both
$\mathbf{n}$ and $\mathcal{M}_i$ are invariant with respect to any
translation, the latter can be disregarded, and we are led to consider only
the magnetic {\it point} group. The presence of an invariant polar vector
$\mathbf{n}$ selects the same subgroup of the bulk magnetic point group as a
homogeneous $\mathbf{E}$ field. It follows that the boundary acquires the
same magnetization components as the bulk with an applied $\mathbf{E}$ field
in the direction of $\mathbf{n}$. Therefore, the boundary develops finite
equilibrium magnetization only if the bulk is magnetoelectric. This
conclusion is equally valid for a metallic AFM whose magnetic point group
would make an insulator magnetoelectric.

\begin{figure}[h]
\includegraphics[width=0.4\textwidth]{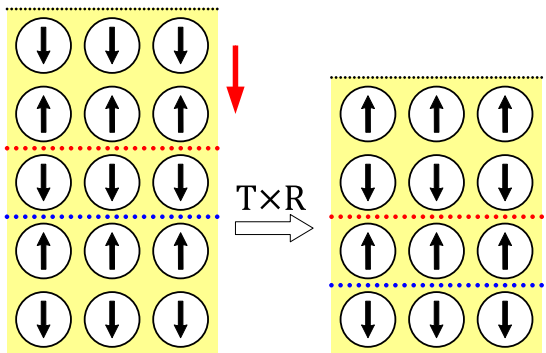}
\caption{Example of a bulk space group operation producing a degenerate
configuration at the surface. An antitranslation $T\times R$, where $T$ is a
pure translation normal to the surface (red arrow) and $R$ is time reversal,
``cuts off'' an atomic layer at the surface of a tetragonal lattice with
layered AFM ordering. This bulk space group operation forbids both
magnetoelectricity and equilibrium boundary magnetization. Black dotted line
shows the boundary. Red and blue dotted lines show two inequivalent types of
boundary/bulk partitioning, which are related by a non-magnetic translation
$T$. The boundary moment averaged over these two partition types vanishes,
because the two types of atomic steps are degenerate.}
\end{figure}

This conclusion reflects the fact that the free energy of the system with a
boundary depends on the polar vector $\mathbf{n}$ as a macroscopic parameter.
Just as in the bulk, the existence of the magnetization at the boundary can
be deduced from the reduction of the bulk magnetic point group by the
presence of the boundary, because translations do not affect $\mathbf{n}$ or
$\mathbf{M}_b$. From a different angle, a boundary can generate a
magnetization only if its zero value is not protected by macroscopic
time-reversal symmetry in the bulk; this singles out the magnetoelectrics.

Thus, equilibrium boundary magnetization of a magnetoelectric is finite
unless $\alpha_{zk}=0$ for all $k$ in the reference frame where $\mathbf{n}$
is parallel to the $z$ axis. If this magnetization is finite for the given
$\mathbf{n}$, it is also necessarily finite for any particular termination
with this orientation, because the magnetic symmetry group of the latter is a
subgroup of the former.

Probes that are not surface-sensitive, such as magnetometry with
$G\{m_i(\mathbf{r})\}=\int m_i(\mathbf{r})d^3r$, measure the sum of
contributions of two film boundaries. The total magnetization is non-zero
unless there is a bulk symmetry operation interchanging the boundaries. It is
always non-zero if these boundaries are with different materials.

The exchange bias induced in a proximate ferromagnetic film by a
magnetoelectric is fundamentally different from the conventional exchange
bias, which is due to a small excess magnetic moment ``frozen-in'' in the AFM
during field-cooling. In particular, this non-equilibrium character typically
leads to an irreversible decline of the exchange bias as the magnetization of
the ferromagnetic layer is repeatedly reversed --- the so-called training
effect \cite{Nogues1,Nogues2}. By contrast, the switchable exchange bias
observed in Ref.\ \onlinecite{Binek} is an equilibrium property and does not
exhibit the training effect.

Since the effect of the boundary is comparable to that of $\mathbf{E}$ of the
crystal-field scale, even simple extrapolation suggests that the induced
magnetic moments at the boundary can be a few orders of magnitude larger than
those achievable due to the bulk magnetoelectric effect. In fact, some
mechanisms do not contain any small parameters and are capable of producing
magnetizations of the order of a few Bohr magnetons per boundary site. I now
classify these mechanisms.

All mechanisms producing linear magnetoelectric response in the bulk
\cite{Fiebig} can generate boundary magnetization as well; these include
changes in (A) the $g$-tensor (here we can also include hybridization-induced
changes of the spin moment), (B) the single-ion anisotropy tensor, (C) the
\emph{intrasublattice} symmetric coupling (including Heisenberg exchange and
dipolar interactions), and (D) the Dzyaloshinsky-Moriya \cite{Dzyal,Moriya}
exchange coupling induced by $\mathbf{E}$ or by the boundary. All of these
except C involve relativistic terms in the Hamiltonian.

Consider the usual case of a collinear AFM. Mechanism C is active only if the
perturbation breaks the equivalence of the AFM sublattices. This symmetry
breaking can be identified by analyzing the so-called black-and-white
(Heesch-Shubnikov) point group based on the decoration of the magnetic sites
with Ising spin variables instead of axial vectors. If the perturbation
(polar vector $\mathbf{E}$ or $\mathbf{n}$) removes all symmetry operations
mapping black and white sites onto each other, the AFM sublattices become
inequivalent; otherwise they do not. If the black-and-white point group
allows Ising ferromagnetism, the true magnetic point group is also
ferromagnetic, but the reverse is not necessarily true. In the first case
mechanism C is allowed, but in the second case the magnetoelectric response
occurs only through spin canting due to a relativistic perturbation. Thus,
certain components of the magnetoelectric tensor, and likewise the boundary
magnetization for certain directions of $\mathbf{n}$, may have no
contribution from mechanism C. For example, consider
Cr$_2$O$_3$ (magnetic point group $\underline{\overline{3}m}$) with
$\mathbf{E}$ or $\mathbf{n}$ oriented parallel to one of the three $U_2$ axes
or to one of the three $\sigma_d$ planes bisecting them. The corresponding
symmetry operation is not removed by $\mathbf{E}$ or $\mathbf{n}$; since both
$U_2$ and $\sigma_d$ interchange the AFM sublattices, the
latter remain equivalent. However, the appearance of $\mathbf{M}$ parallel to
$\mathbf{E}$ or $\mathbf{M}_b$ parallel to $\mathbf{n}$ through spin canting
is allowed.

\begin{figure*}[htb]
\includegraphics[width=0.85\textwidth]{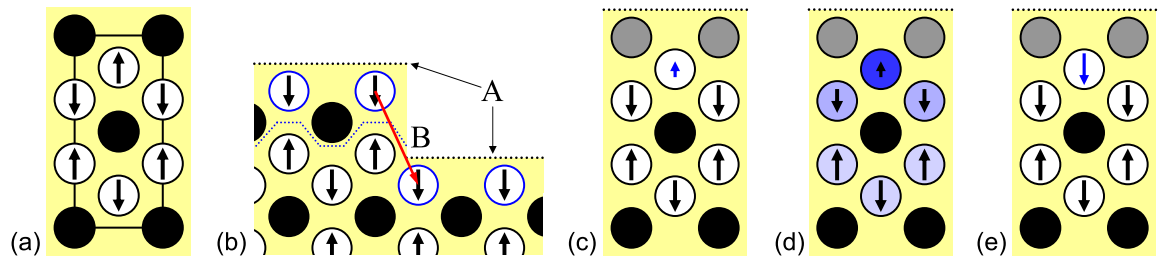}
\caption{(a): Projected unit cell of the trirutile lattice assumed by
magnetoelectric Fe$_2$TeO$_6$ (schematic) \cite{Kunnmann}. Black circles: Te
atoms; circles with arrows: Fe atoms with spin directions. O atoms, whose
positions make the lattice non-symmorphic, are not shown. (b): Boundary
termination cutting between the Fe layers (indicated as type A), including an
atomic step. The same AFM sublattice appears at the surface (blue circles).
The red arrow shows the bulk symmetry operation (4-fold screw rotation)
mapping degenerate atomic steps onto each other. Termination type B (blue
dotted line) puts another sublattice at the surface, but it is not related by
symmetry and not degenerate with type A. (c)-(e): Mechanisms affecting the
boundary magnetization. Termination type B is used as example; Te sites shown
in gray may or may not be occupied. (c) Mechanism S1: Changed spin state
(blue arrow). (d) Mechanism S2: Unequal thermal averages. Shades of blue and
arrow size indicate the degree of thermal disorder. (e) Mechanism S3: Flipped
spin direction (blue arrow).}
\end{figure*}

If the equivalence of the AFM sublattices \emph{is} broken by the boundary,
the consequences are far more drastic than in the bulk mechanism C. For any
particular boundary termination, even without reconstruction, the sites
corresponding to different AFM sublattices are structurally different. For
example, all the sites closest to the boundary can have spins ``down,'' while
there is no equivalent termination with boundary spins ``up.'' The situation
is illustrated in Fig.\ 2b using Fe$_2$TeO$_6$ (magnetic point group
$4/\underline{mmm}$) as an example. In this figure, terminations A (with
boundary spins up) and B (with boundary spins down) are structurally
distinct, and therefore they occur with different probabilities in
equilibrium. Even if they did appear with equal weights, they are
inequivalent magnetically, and don't generally add up to zero. Indeed, there
are several mechanisms leading to the deviation of the average magnetic
moments on the boundary sites from the bulk ones (see Fig.\ 2c-2e): (S1)
Different local environment of the magnetic sites near the boundary leads to
a different local magnetic moment, and perhaps even a different atomic
multiplet. (S2) Since the translational symmetry is broken by the boundary,
any symmetric exchange interaction (even purely \emph{intersublattice} one)
leads to different thermal averages at the boundary. (S3) The exchange
coupling near the boundary can be so different from the bulk that the AFM
ordering pattern can change to ferrimagnetic there. Mechanism S1 can be
viewed as a boundary analog of bulk mechanism A, and S2 is the boundary
counterpart of mechanism C. S1 and S2 are always present if the black-white
symmetry is broken. Apart from these mechanisms affecting the magnitudes of
the local moments, the \emph{coupling} of these moments to the external probe
can be different. For example, in the exchange bias setup, the sites closest
to the boundary are expected to have the strongest exchange coupling to the
proximate ferromagnet.

The bulk linear magnetoelectric effect appears in the free energy expansion
as a second-rank tensor $\alpha_{ik}$. This is not the case for the
equilibrium boundary magnetization. Since the free energy is a non-analytic
function of $\mathbf{n}$ \cite{LL5}, the boundary magnetization, given by its
field derivative, is also non-analytic. Thus, even if the AFM sublattices are
equivalent for some symmetric directions of $\mathbf{n}$, forbidding
mechanisms S1-S3 for these orientations, these mechanisms can still
generate large boundary magnetization for less symmetric orientations.

Boundary magnetization $\mathbf{M}_b$ vanishes at the bulk N\'eel
temperature, but different mechanisms can be partially distinguished
experimentally based on its temperature dependence. The thermal mechanism S2
can lead to a non-monotonic contribution with a maximum, similar to the bulk
behavior of mechanism C (cf.\ $\alpha_{zz}$ in Cr$_2$O$_3$). Other mechanisms
should lead to $\mathbf{M}_b$ monotonically decreasing with $T$. All these
mechanisms do not contain any small parameters and can produce $\mathbf{M}_b$
of the order of a few Bohr magnetons per boundary
site. The non-monotonic temperature dependence of the exchange bias field
observed in the heterostructure of Ref.\ \cite{Binek} suggests that mechanism
S2 plays an important role at the Cr$_2$O$_3$(0001)/Pd interface.

Roughness-insensitive mechanisms based on Dzyaloshinskii-Moriya interaction
at the compensated interface were proposed \cite{Dagotto} to explain the
exchange bias induced by single-domain multiferroic BiFeO$_3$ in a proximate
FM \cite{Ramesh}. They are, however, fundamentally different from all the
mechanisms discussed here, because BiFeO$_3$ is weakly ferromagnetic
in the bulk, and also its $\mathbf{M}$ and $\mathbf{P}$ are not linearly
coupled \cite{Spaldin}. Therefore, the boundary does not induce additional
magnetization components that are forbidden in the bulk.

Much attention is devoted to the search of a room-temperature multiferroic
material with linear coupling between the electric polarization $\mathbf{P}$
and magnetization $\mathbf{M}$, because its $\mathbf{M}$ could be switched
along with $\mathbf{P}$ by electric field only
\cite{Eerenstein,RS,Cheong,Bibes,Fennie}. The paraelectric phase of such a
material is a magnetoelectric AFM \cite{Scott,Fennie}. The desirable geometry
involves voltage applied across the multiferroic film, with $\mathbf{E}$
normal to its surface \cite{Bibes}. Equilibrium magnetization $\mathbf{M}_b$
necessarily exists at the boundary of such a material. This $\mathbf{M}_b$
has the same components as the bulk $\mathbf{M}$ coupled to $\mathbf{P}$, but
$\mathbf{M}_b$ is coupled to the bulk AFM order parameter. Ferroelectric
switching directly switches only the part of $\mathbf{M}$ linearly coupled to
$\mathbf{P}$, but not $\mathbf{M}_b$. In addition, the states with parallel
or antiparallel $\mathbf{M}_b$ and $\mathbf{M}$ are non-degenerate, meaning
that one of them is metastable or even unstable. Thus, the presence of
intrinsic boundary magnetization may hamper purely electric magnetization
switching using a multiferroic with linear coupling of $\mathbf{P}$ and
$\mathbf{M}$.

Equilibrium boundary magnetization of magnetoelectric AFMs has far-reaching
consequences for the design of magnetic nanostructures. First, very large
exchange bias fields should be achievable in magnetoelectric/ferromagnet
bilayers, comparable to the estimates for a fully uncompensated AFM interface
\cite{Meiklejohn,Kiwi}. Second, magnetoelectric AFMs are precisely the
materials that can be switched between the time-reversed AFM domain variants
by a simultaneous application of $\mathbf{E}$ and $\mathbf{B}$ fields
\cite{Schmid,Martin}, thereby switching the boundary magnetization and the
exchange bias field \cite{Binek}. The $\mathbf{B}$ field may be permanent,
while $\mathbf{E}$ may be provided by a voltage pulse across the
magnetoelectric film. Since no depolarizing fields or elastic strains are
involved, the AFM domains are stable, and this switching is fully
non-volatile. Some device architectures based on the magnetoelectric active
layer, where the operation is based on the linear bulk magnetoelectric
response, were described by Binek and Doudin \cite{BD}. These architectures
are greatly facilitated by the existence of a switchable equilibrium boundary
magnetization, which moreover allows the AFM domain state to be used as a
switchable state variable.

In summary, symmetry requires that magnetoelectric antiferromagnets
possess a finite boundary magnetization in thermodynamic equilibrium. This
magnetization is particularly large if the boundary breaks the equivalence of
the antiferromagnetic sublattices; specific microscopic mechanisms have been
classified. This understanding will hopefully stimulate further studies of
boundary magnetization of magnetoelectrics and its exploitation in
nanoelectronic devices.

This work was supported by NSF MRSEC, the Nanoelectronics Research
Initiative, and Nebraska Research Initiative. The author is a Cottrell
Scholar of Research Corporation.


\begin{thebibliography}{29}
\bibitem[{not()}]{note-staggered}
\bibinfo{note}{Formally, the thermodynamic limit should be taken with a
  staggered magnetic field applied at the magnetic sites, whose magnitude is
  sent to zero afterwards.}
\bibitem[{pro()}]{probe-note}
\bibinfo{note}{A spin-selective probe with a lower symmetry can generate a
  finite spin signal even from a magnetically compensated boundary, but we wish
  to reveal the intrinsic spin asymmetry of the boundary.}
\end{thebibliography}
\end{document}